\documentclass[twocolumn,showpacs,preprintnumbers,amsmath,amssymb]{revtex4}
\usepackage{graphicx}
\usepackage{dcolumn}
\usepackage{color}
 
\newcommand{\beq}{\begin{equation}}
\newcommand{\eeq}{\end{equation}}

\begin{document}

\title{Anisotropic fluxes and nonlocal interactions in MHD turbulence}

\author{A. Alexakis, B. Bigot, H. Politano}
\affiliation{Laboratoire Cassiop\'ee, UMR 6202,
             Observatoire de la C\^ote d'Azur,
             BP 4229, Nice Cedex 4, France}

\begin{abstract}

{We investigate the locality or nonlocality of the energy transfer
and of the spectral interactions involved in the cascade for decaying 
magnetohydrodynamic (MHD) flows in the presence of a uniform magnetic field
$\bf B$ at various intensities.
The results are based on a detailed analysis of
three-dimensional numerical flows at moderate Reynold numbers.
The energy transfer functions, as well as the global and partial fluxes,
are examined by means of different geometrical wavenumber shells.
On the one hand, the transfer functions of the two conserved Els\"asser energies $E^+$ and $E^-$
are found local in both the directions parallel ($k_\|$-direction)
and perpendicular ($k_\perp$-direction) to the magnetic guide-field,
whatever the ${\bf B}$-strength.
On the other hand, from the flux analysis, 
the interactions between the two counterpropagating Els\"asser waves
become nonlocal.
Indeed, as the ${\bf B}$-intensity is increased,
local interactions are strongly decreased and
the interactions with small $k_\|$ modes dominate the cascade.
Most of the energy flux in the  $k_\perp$-direction is due to modes in the plane at  $k_\|=0$,
while the weaker cascade in the $k_\|$-direction is due to the modes with
$k_\|=1$. The stronger magnetized flows tends thus to get closer to the weak turbulence
limit where the three-wave resonant interactions are dominating.
Hence, the transition from the strong to the weak turbulence regime
occurs by reducing the number of effective modes in the energy cascade.
}
\end{abstract}


\maketitle  

\section{Introduction}

The existence of magnetic fields is known in many astrophysical
objects such as interstellar medium, galaxies, accretion discs, star and
planet interiors, or solar wind (see e.g. \cite{Zeldovich}). 
In most of these systems, the magnetic fields
are strong enough to play a significant dynamical role.
The involved kinetic and magnetic Reynolds
numbers in these astrophysical bodies are large enough so that
the flows exhibit a turbulent behavior with a large
continuous range of excited scales, from the largest where energy is injected towards the 
finest where energy is dissipated.
In many cases, a strong large-scale magnetic field is present and induces dynamic anisotropy.
Direct numerical simulations that examine in detail the turbulent processes
in geo- and astrophysical plasmas are very
difficult to achieve, for only rather modest Reynolds numbers can be reached
with nowadays computers. One way around this difficulty is to model 
the small spatial and temporal scales to reproduce the large-scale behavior of
turbulent flows. A more basic understanding of turbulence is thus needed 
to adequately model the flows, in particular when a uniform magnetic
field, constant both in space and time, is applied. 

As a first approximation, the incompressible magnetohydrodynamic
(MHD) equations can be used to describe the evolution of both
velocity, ${\bf u}$, and magnetic field, ${\bf b}$, fluctuations. 
In the presence of an uniform magnetic field ${\bf B}$ (magnetic fields are here expressed in velocity units), 
the Els\"asser
formulation of the MHD equations, with constant unit mass density,
reads
\begin{equation}
\label{MHD}
\partial_t {\bf z}^\pm = \pm {\bf B} \cdot \nabla {\bf z}^\pm 
- {\bf z}^\mp  \cdot \nabla {\bf z}^\pm -\nabla P + \nu \nabla^2 {\bf z}^\pm
\end{equation}
together with $\nabla \cdot {\bf z}^\pm=0$, where
${\bf z}^\pm={\bf u\pm b}$ are the  Els\"asser fluctuations, and
$P$ is the total (kinetic plus magnetic) pressure.
We assume here equal molecular viscosity
$\nu$ and  magnetic diffusivity $\eta$, in other words a unit magnetic Prandtl number ($P_r=\nu/\eta=1$). 
Hereafter, the direction along the ${\bf B}$ magnetic field
is referred to as the parallel direction and the projection of the wavevectors along this
direction is denoted $k_\|$, while the two directions of planes perpendicular to ${\bf B}$
are referred to as perpendicular directions, the wavevector projection onto such planes
being denoted ${\bf k}_\perp$ with norm $ k_\perp \equiv |{\bf k}_\perp|$. 

For periodic boundary conditions, equations (\ref{MHD}) have two independent invariants 
in the absence of molecular  viscosity and magnetic diffusivity, namely the Els\"asser energies : 
\beq
\label{InvE+-}
E^{\pm}=\frac{1}{2}\int {{\bf z}^\pm}^2({\bf x}) \ dx^3.
\eeq
However, when very small viscosity and magnetic diffusivity are present, it is 
expected that the nonlinear terms cascade the energies between scales, in
the so-called inertial range, up to smallest ones
where dissipation becomes effective and removes energy from the system. 
The rate at which large scales lose
energy is then controlled by the nonlinear terms ${\bf z}^\mp  \cdot
\nabla {\bf z}^\pm -\nabla P$ that are responsible for coupling
different scales and cascading the energy towards smaller and smaller
scales. The nature of the interactions among various scales in
turbulent flows that lead to this cascade is a long standing problem. 
Understanding the involved mechanisms 
is very important to predict evolution of the large-scale flow behavior,
and to estimate global quantities in astrophysical systems, such as the transport of
angular momentum, and accretion rates in accretion discs.

High-Reynolds-number hydrodynamic
turbulence, often investigated in the framework of
statistically homogeneous and isotropic turbulence (which can be
questionable in natural flows), is described to first order by
the Kolmogorov theory \cite{Kolmogorov1941}. In this phenomenological description, 
interactions between eddies of similar size give the dominant contribution
to the energy cascade. This assumption leads to an 
energy spectrum in $k^{-5/3}$ and an energy cascade rate
proportional to $u_{rms}^3/L$, where $u_{rms}$ is the root-mean-square
of the velocity at large scale and $L$ is the typical (large) flow scale.

The cascade in MHD turbulence is more complex, especially in the presence
of a background magnetic field.
Even in the simplest case of zero or small intensity of the ${\bf B}$-field,
so that isotropy could be recovered, 
whether the MHD energy cascade can be described by a phenomenology {\it \` a la}
Kolmogorov is still an open question. 
{In particular, the assumption that interactions between similar size eddies 
(local interactions) are responsible for the cascade of energy to smaller scales 
has been questioned in turbulent MHD flows 
both  by theoretical arguments \cite{Vermarev04,Verma2005,Yousef2007} and the use of numerical simulations 
\cite{Alexakis2005,Mininni2005,Carati2006}.
It has been shown for mechanically forced MHD turbulence
there is a strong nonlocal coupling between the 
forced scales and the small scales of the inertial range.}
Moreover, the large-scale magnetic field
generated by the dynamo action can also locally affect the small scales
by suppressing the cascade rate
in the same manner that an initially imposed uniform magnetic field would.
In the other limit, a strong ${\bf B}$-field 
can lead to the flow bi-dimensionnalization, with a drastic reduction of
the nonlinear transfers along the uniform magnetic field. 
For a ${\bf B}$-intensity (denoted $B$) well above the {\it rms} level of kinetic and magnetic fluctuations,
the MHD turbulence may be dominated by the Alfv\'en waves dynamics, 
leading to wave (or weak) turbulence 
where the energy transfer, stemming from three-wave resonant interactions,
can only increase perpendicular components of the wavevectors, {\it i.e.} components in planes perpendicular to
the ${\bf B}$-direction ($k_{\perp}$-direction), the nonlinear transfers along ${\bf B}$
($k_\|$-direction)
being completely inhibited \cite{Galtier2000,Galtier2002}. 
How MHD turbulence moves from the weak turbulence limit,
$B \gg u_{rms}$, to the strong turbulence limit, $B \sim u_{rms}$ and $B \sim 0$ (where isotropy
could be recovered), is an open question.

Various authors have tried  to give a physical description of the
strong turbulence regime with $B \sim u_{rms}$. 
Iroshnikov \cite{Iroshnikov1963} and Kraichnan \cite{Kraichnan1965} 
first proposed a phenomenological description that
takes into account the effect of a large-scale magnetic field
by reducing the rate of
the cascade due to the short time duration of individual collisions of
$z^{\pm}$ wave packets. 
%
The resulting 1-D energy spectrum is then given by 
$E(k)\sim (\epsilon B)^{1/3} k^{-3/2}$. 
However, this description assumes isotropy and,
while the effect of the large-scale field is taken into
account by reducing the effective amplitude of the interactions, the
interactions themselves are considered to be local. 
In order to take into account
anisotropy in strong turbulence, a scale dependent anisotropy has been
proposed \citep{Goldreich1995}, 
the turbulent $z^{\pm}_l$-eddies being such that the associated Alfv\'en  
$\tau_{_A}\sim l_\|/B$ and  nonlinear
$\tau_{_{NL}}\sim l_\perp/z $ times are equal (the so called critical balance),
where $l_\|$ and $l_\perp$ are the typical length scales respectively parallel and
perpendicular to the mean magnetic field.
Repeating the Kolmogorov arguments, one ends up with a
$E(k_\|,k_\perp)\sim k_\perp^{-5/3}$ energy spectrum with 
$k_\|\sim k_\perp^{2/3}$.
%
%
Recently, this result has been generalized  
in an attempt to model MHD 
turbulence both in the weak and the strong limit,
the ratio of the two time scales $\tau_{_A}/\tau_{_{NL}}$ being
kept fixed but not necessarily of order one \cite{Galtier2005}. 
{In an other approach to obtain the transition from the
strong to the weak turbulence limit \citep{Matthaeus1989,Zhou2004} suggested
time scale for the energy cascade is given by
the inverse average between the Alfv\'en and the nonlinear time scale
$\tau^{-1}=\tau_{_A}^{-1}+\tau_{_{NL}}^{-1}$.}
All these models however assume locality of interactions that 
are also in question in anisotropic MHD turbulence \cite{Bhatta2001}.     
A nonlocal model for anisotropic turbulence
has been recently proposed by one of the authors \cite{Alexakis2007};
it assumes that the energy cascade
is due to interactions between eddies 
with different parallel sizes and similar perpendicular scales,
while a non-universal behavior is expected for moderate Reynold numbers. 

Although very useful in getting a first order understanding
of the processes involved in a turbulent cascade, cascade-energy models have to be unavoidably
based on assumptions that need to be tested. To this respect, numerical simulations
of the MHD equations are very valuable because they provide information about the evolution of
the fields in the whole space, something not easily obtained from observations.
Many numerical investigations have been performed   
during the last two decades 
\citep{Mason2007,Muller2005,Ng2003,Dmitruk2003,Maron2001,Cho2000,Biskamp2000,Ng1996,Oughton94}
and, at the achieved Reynolds numbers, they have demonstrated that 
different power-law exponents are obtained depending on the chosen forcing.
In this work, we use the results of numerical simulations 
of free decaying MHD flows at moderate Reynolds number to
investigate the MHD interactions for various intensities of
the external magnetic field.
In particular, we try to investigate 
whether the transfer of energy in the parallel and perpendicular direction is local
(i.e. the two energies $E^\pm$ cascade between nearby wavenumbers) or nonlocal 
(i.e. distant wavenumbers are involved in the cascade),
and whether the coupling between the two oppositely moving waves ${\bf z}^+$ and ${\bf z}^-$
(that do not exchange energy) is local or not; and if not, which modes are responsible for 
the energy cascade.

The paper is organized as follows. In the next section,
we give the precise definitions of the transfer functions and partial fluxes
used to analyze the nature of the energy cascade.
The details of the numerical simulations are given in Section IIIa.
In Section IIIb, we investigate the locality or nonlocality of the energy transfers,
and in section IIIc, we examine the nature of the interactions
between the two ${\bf z}^+$ and ${\bf z}^-$ fields.
We conclude and discuss our results in Section IV.

\section{Definitions}

Our goal is to investigate the interactions among different scales.
To define the notion of ``scale",
we use the field Fourier transforms :
\beq
\label{Fourier}
  {\bf \widehat{z}^{\pm}(k) } = \frac{1}{({2\pi})^3}\int {\bf {z^{\pm}}(x)}e^{-i{\bf k\cdot x}} dx^3 
\eeq
defined in a $2\pi$-periodic cube, such as 
\beq
\label{Space}
{\bf  z^\pm(x)}           = \sum_{\bf k} {\bf \widehat{z}^{\pm}}({\bf k})e^{i{\bf k\cdot x}}, 
\eeq
Similar size eddies will be considered as the ones whose Fourier transform
contains similar wavenumbers.

In any basic flow interaction, three wavevectors are involved.
For example, the evolution of 
a given Fourier amplitude $\widehat{\bf z}^{+}({\bf k})$ 
will be coupled to a $\widehat{\bf z}^{-}({\bf p})$ one and cascade the energy
to the mode $\widehat{\bf z}^{+}({\bf q})$ such that the wavevectors satisfy ${\bf k+p+q}=0$.
Note that the mode $\widehat{\bf z}^{-}({\bf p})$ does not gain or lose energy
from this interaction since the two energies $E^+$ and $E^-$ are separately
conserved. To obtain the cascade mean rate, 
one needs to average over all possible triadic interactions.
To get a phenomenological understanding of the processes
at play in MHD turbulence, we need to know if : 
i) most of the energetic exchanges occur between wavenumbers such that
$|{\bf k}|\sim |{\bf q}|$ and ii) the energy flux is a result of spectral
interactions of the two fields ${\bf z}^\pm$ with similar wavenumbers or not
{($|{\bf k}|\sim |{\bf p}|$)}.

To address these questions, let us 
consider a partition of the wavevectors into non-overlapping sets $S^\pm_{_K}$
such that $S^\pm=\bigcup_{K=1}^{\infty}S^\pm_{_K}=\mathbb{Z}^3$.
For example $S^\pm_{_K}$ could be the spherical shells of unit width and radius $K$,
{\it i.e.} set of wavevectors {\bf k} that have $K<|{\bf k}|\le K+1$.
We now define the filtered fields ${\bf z}^\pm_{_K}({\bf x})$
so that only modes in the set $S^\pm_{_K}$ are kept:
\beq
\label{filterz}
{\bf       z^\pm_{_K}(x)} = \sum_{{\bf k} \in S^\pm_{_K}} {\bf \widehat{z}^{\pm}}({\bf k})e^{i{\bf k \cdot x}}.
\eeq
Clearly, one gets
\beq
{\bf       z^\pm(x) = \sum_{K}  z^\pm_{_K}(x)}. 
\eeq

The triadic interactions among the different sets, say $S^\pm_{_K}$,$S^\mp_{_P}$ and $S^\pm_{_Q}$, are given by:
\beq
\label{triadic}
\mathcal{T}^\pm_3(K,P,Q)=-{\int {\bf z^\pm_{_K}} {\bf z^\mp_{_P}} \cdot \nabla {\bf z^\pm_{_Q}}dx^3}
\eeq
that express the rates at which $E^\pm$ energies are transfered from $S^\pm_{_Q}$ to 
$S^\pm_{_K}$ sets due to the interactions with the modes belonging to $S^\mp_{_P}$ set.
Note that the collection of sets $S^+$ and $S^-$ need not to be necessarily the same;
for example, $S^+$ could be a collection of cylindrical shells while $S^-$ 
could be a collection of plane sheets.
Adding over the index $P$ (all sets in $S^\mp$), we obtain the transfer functions :
\beq
\label{transfer}
\mathcal{T}^\pm(K,Q)=\sum_P \mathcal{T}^\pm_3(K,P,Q)=
                      -{\int {\bf z^\pm_{_K}} {\bf z^\mp} \cdot \nabla {\bf z^\pm_{_Q}}dx^3}
\eeq
that give the $E^+$ and $E^-$ transfer rates from $S^\pm_{_Q}$ to 
$S^\pm_{_K}$ sets due to all possible interactions. 
Note that the ${\bf z}^+$ field is not exchanging energy with
the ${\bf z}^-$ field, and vice versa, but their interaction is responsible 
for the redistribution of the energy among various sets.
$\mathcal{T}^\pm(K,Q)$ can give us information about the locality or nonlocality of the energy transfer,
{\it i.e.} whether the energy is exchanged by nearby sets or long-range transfers from the large
scales directly to the small scales are also involved. 

However, the $\mathcal{T}^\pm(K,Q)$ transfer functions 
do not give us direct information on the scales of the two fields ${\bf z}^+$ and ${\bf z}^-$ that interact
and contribute to the energy cascade. 
To investigate the locality on nonlocality of the interactions between 
the two Els\"asser counterpropagating waves, we introduce the partial fluxes 
(see \cite{Alexakis2005b,Mininni2006})
defined as:
\begin{eqnarray}
\Pi^\pm_{_P}(K)=&  \sum_{K'=0}^K \sum_{Q=0}^\infty \mathcal{T}^\pm_3(K',P,Q) \nonumber \\
               =& -\sum_{K'=0}^K\int {\bf z^\pm_{_{K'}} {\bf z}_{_P}^\mp \cdot \nabla {\bf z}^{\pm}} dx^3
\label{partialflux}
\end{eqnarray}
that express the flux of energy out of the outer surface of the $S^\pm_{_K}$ shell due to the interactions
with the $S^\mp_P$ shell. 
Summation over the whole $S^\mp$ collection of sets enable to recover the usual definition for the global fluxes :
\begin{eqnarray}
\Pi^\pm(K)=&  \sum_{K'=0}^K \sum_{Q=0}^\infty \sum_{P=0}^\infty \mathcal{T}^\pm_3(K',P,Q)  \nonumber \\
          =&       -\sum_{K'=0}^K\int {\bf z^\pm_{_{K'}} {\bf z}^\mp \cdot \nabla {\bf z}^{\pm}} dx^3
\label{flux}
\end{eqnarray}
 
In the current work, we are going to use three different types of wavevector collections.
We first consider spherical shells
traditionally used in studies of isotropic turbulence so
that a set $S_{_K}$ contains the wavevectors $\bf k$ such that 
$K\le {|\bf k|} < K+1$. 
The second collection of sets are cylindrical
shells along the direction of the guiding magnetic field. 
In this case, the set $S_{_K}$ contains the wavevectors $\bf k$ such that 
$K\le k_{\perp} < K+1$ (with $k_{\perp}=\sqrt{k_x^2 +k_y^2}$). 
Finally, we consider planes perpendicular to the
${\bf B}$-direction, so that the set $S_{_K}$ 
contains the wavevectors $\bf k$ whose $k_\|$-component satisfies 
$K\le |k_\|| <K+1$ (where $k_\|$ stands for $k_z$).

\section{Numerical results}
\subsection{Numerical setup and initial conditions}
We integrate numerically the three-dimensional incompressible MHD equations
(\ref{MHD}), in a $2\pi$-periodic box using a pseudo-spectral method
with $256^3$ collocation points. The time marching uses an
Adams-Bashforth Cranck-Nicholson scheme, i.e. a second-order
finite-difference scheme. The initial kinetic and magnetic fields
correspond to spectra proportional to $k^2 exp(-k/2)^2$ for
$k=[1,8]$, which means a flat modal spectrum for wavevector ${\bf
k}$ up to $k=2$, to prevent any favored wavevector at time $t=0$, and
the associated kinetic and magnetic energies are chosen equal, namely
$E_v(t=0)=E_b(t=0)=1/2$, as in previous numerical studies (see
\cite{Milano} and references therein). 
{Moreover, the correlation between the velocity and magnetic field fluctuations, as measured by
the cross-correlation coefficient defined by $2 \int {\bf v}({\bf x}) \cdot {\bf b}({\bf x)} \ dx^3/(E_v+E_b)$, 
is initially less than $1\%$.}
At scale injection, the
initial kinetic and magnetic Reynolds numbers are about $800$ for
flows at $\nu=\eta \sim 4 \times 10^{-3}$, with $u_{rms}=b_{rms}=1$
and an isotropic integral scale $L = 2\pi \int{k^{-1} E_v(k) dk}/
\int{E_v(k) dk}$ of about $\pi$. The dynamics of the flow is then let to freely
evolve. The parametric study according to the intensity of the
background magnetic field ${\bf B}$ is performed for four different
values : $B = 0, 1, 5$ and $15$. 
All the simulations are run up to a computational time $t_{max}=15$, 
at which the loss of the total energy (kinetic plus magnetic) is about $95\%$ for the simulation with $B=0$,
$90\%$ for $B=1$, and $83\%$ for the $B=5$ and $B=15$ runs.

\begin{figure} 
\includegraphics[width=8cm]{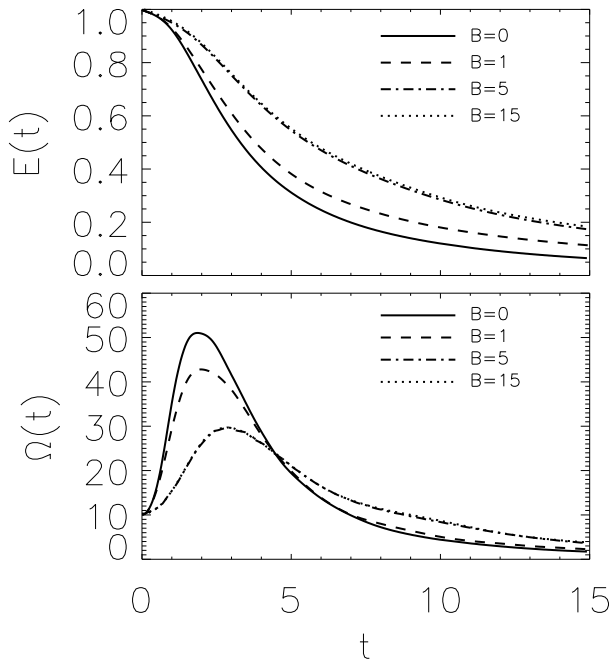}
\caption{Temporal evolutions of the total energy (top panel)  and of 
the total enstrophy (bottom panel) for the four examined intensities of 
the ${\bf B}$ applied field.
\label{fig1}}
\end{figure}

\begin{figure}
\includegraphics[width=8cm]{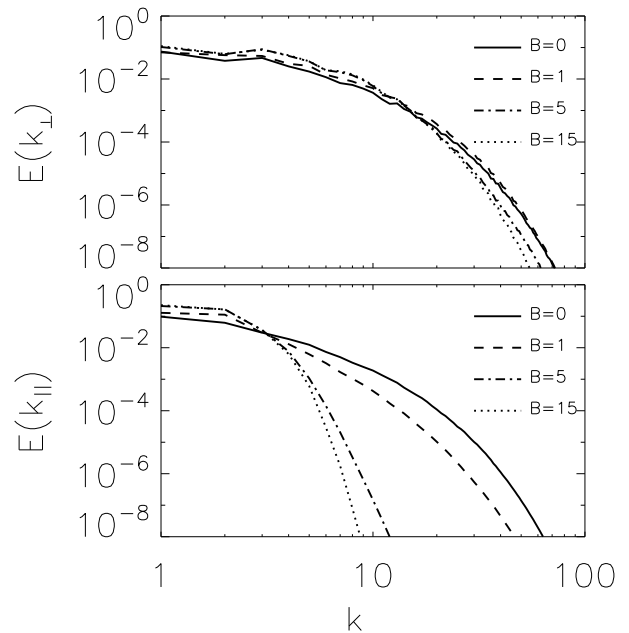}
\caption{Total energy spectra in the perpendicular (top panel)
and the parallel (bottom) directions.
\label{fig2}}
\end{figure}

Figure \ref{fig1} shows the time evolution of the total energy, $E(t)=(E^+(t) +E^-(t))/2$,
and  the total enstrophy, 
$\Omega(t)=1/2 \int [{{\bf w}}^2({\bf x},t)+ {{\bf j}}^2({\bf x},t)] \ dx^3$
(where ${\bf w}=\nabla \times {\bf u}$ stands for the vorticity field,
and ${\bf j}= \nabla \times {\bf b}$ for the current),
for the four different simulations. 
{One can note that the influence of the strength of the external magnetic field 
clearly slows down the flow dynamics}.
Note that  the $B=5$ and $B=15$ flows present a very similar temporal behavior
in energy as well as in enstrophy. 
{
The analysis  that follows in the next subsection is based 
on the outputs of the runs at $t = 4$ where
the spectra are fully developed and all runs wave roughly the same enstrophy.
}
{At this time, the cross-correlation coefficient, that one can also write 
$(E^+-E^-)/(E^++E^-)$, is about $3.6\%$ 
and $2.5\%$ respectively for the $B=0$ and $B=1$ flows, while it is 
close to $1.6\%$ for $B=5$ and $1\%$ for $B=15$, with thus a lesser increase 
in the stronger magnetized flows.}
The energy spectra 
in the perpendicular direction to the uniform ${\bf B}$-field 
$E(k_\perp)=1/2 \int [\hat{\bf z}^-(k)]^2+[\hat{\bf z}^+(k)]^2 k_\perp dk_\|   $
and in the parallel one  
$E(k_\|)=1/2 \int [\hat{\bf z}^-(k)]^2+[\hat{\bf z}^+(k)]^2 k_\perp dk_\perp$
are shown in Figure \ref{fig2} at the same time.
Clearly as the magnetic field intensity is increased, the spectrum in the 
$k_\|$-direction becomes steeper. 
Because the planes at $k_\|=0$ and $k_\|=1$ are shown to play an important role
in the cascade, we mention here their properties in more details.
In absence of the applied magnetic field, the modes with $k_\|=0$ contain $32\%$ of 
the total energy and $8\%$ of the total enstrophy, and the $k_\|=1$ modes have 
$30\%$ of the total energy and $10\%$ of the total enstrophy.
In the strongly anisotropic case, $B=15$,
the $k_\|=0$ modes contain $55\%$ of the total energy and
$34\%$ of the total enstrophy while the $k_\|=1$ modes have
$37\%$ of the total energy and $35\%$ of the total enstrophy.

\begin{figure}
\includegraphics[width=8cm]{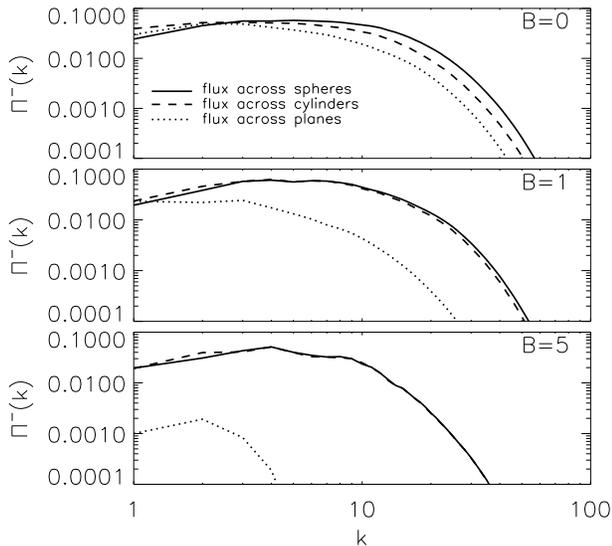}
\caption{Fluxes $\Pi_S^{-}(k)$ across : (i) spheres (solid line),
(ii) cylinders (dashed line), (iii) planes (doted line), for $B=0$ (top panel),
$B=1$ (mid panel) and $B=5$ (bottom panel).
\label{fig3}}
\end{figure}

In our investigation, we focus on the cascade
of the $E^-$ energy. The $E^+$ cascade has also been analyzed
and it gives qualitatively similar results. 
We consider separately
the cascades in the perpendicular and parallel directions relatively to the applied 
magnetic field.
For this reason, we examine three different types of flux; 
(i) the flux across spheres of radius $k \equiv |{\bf k}|$ that
corresponds to an isotropic analysis, 
(ii) the flux across cylinders of radius $k \equiv k_{\perp}$ that
corresponds to the flux in the perpendicular direction 
and (iii) the flux across planes located at $k \equiv |k_\||$ that corresponds to the flux 
in the direction parallel to the ${\bf B}$-field direction.
Figure \ref{fig3} shows these three fluxes, as a function of $k$, 
for various ${\bf B}$-intensities.
It is clear that as the amplitude of the large-scale magnetic field is increased,
the parallel flux is strongly reduced. 
For $B=5$, this flux is reduced by more than one order of magnitude 
when compared to the case with $B=0$. 
For $B=15$, the parallel flux across planes is very small and 
it even takes negative values.

\subsection{Energy transfers}

\begin{figure}
\includegraphics[width=8cm]{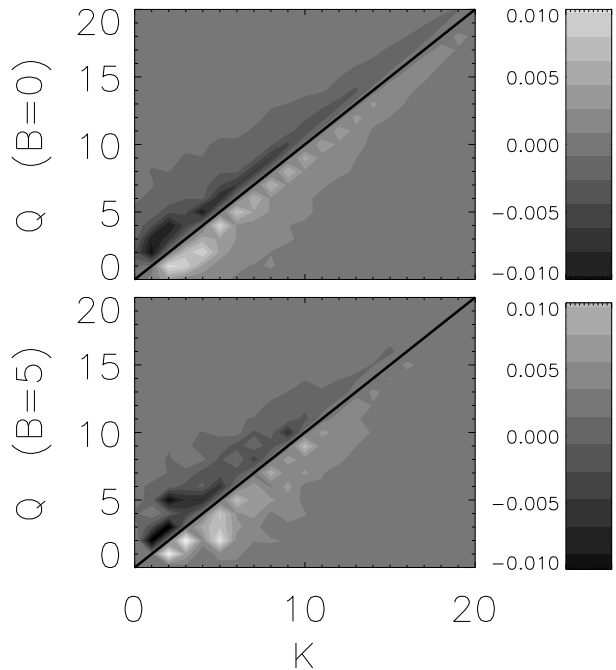}
\caption{The transfer function $\mathcal{T}^-(K,Q)$
that demonstrates the energy exchange between {\it cylindrical shells}
of radius $K$ and $Q$. Solid lines show the diagonal $K=Q$.
The top panel shows the $B=0$ case and the bottom panel the $B=5$ case. 
\label{fig4}}
\end{figure}

\begin{figure}
\includegraphics[width=8cm]{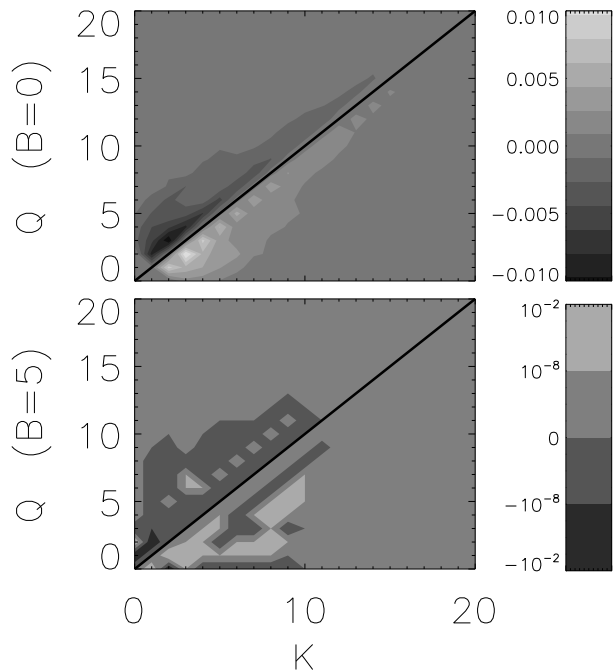}
\caption{The transfer function $\mathcal{T}^-(K,Q)$
that demonstrates energy exchanges between {\it plane sheets}
located at distance $K$ and $Q$ from origin, for
$B=0$ (top panel) and $B=5$ (bottom panel) cases.
Solid lines indicate the $K=Q$ diagonal.
\label{fig5}}
\end{figure}

We now examine the locality or nonlocality
of energy transfers from our numerical data.
For two different values of the uniform magnetic field, namely $B=0$ and $B=5$, 
Figure \ref{fig4} shows a shadow-graph of the transfer function $\mathcal{T}^-(K,Q)$ between
${\bf z^-_K}$ and ${\bf z^-_Q}$, defined in Eqs. (\ref{filterz}) and (\ref{transfer}),
for energy exchanges across cylindrical shells (perpendicular cascade),  
while Figure \ref{fig5} shows the transfer function $\mathcal{T}^-(K,Q)$
for energy exchanges across plane sheets (parallel cascade). 
In all cases, the transfer is concentrated along the diagonal
$K=Q$ line. This indicates that the cascade happens through a local energy exchange.
Similar results are obtained from the two other simulations at $B=1$ and $B=15$ (not shown).
Note the highly non-linear color bar used for the parallel cascade in the $B=5$ case.
This choice is due to the extremely fast decrease of the amplitude of $\mathcal{T}^-(K,Q)$
as the wavenumbers $K$ and $Q$ become large. 
From Figures \ref{fig4} and \ref{fig5}, it can be seen that most of the energy exchange 
happens close to the diagonal line ($K=Q$). This implies that waves traveling in the same direction
exchange energy between similar size wavenumbers.
In the strong $B$ flow, some inverse cascade is also visible in the parallel cascade 
(Fig. \ref{fig5}) as 
indicated by the dark lines below the diagonal and the bright ones above the diagonal. 

\begin{figure}
\includegraphics[width=8cm]{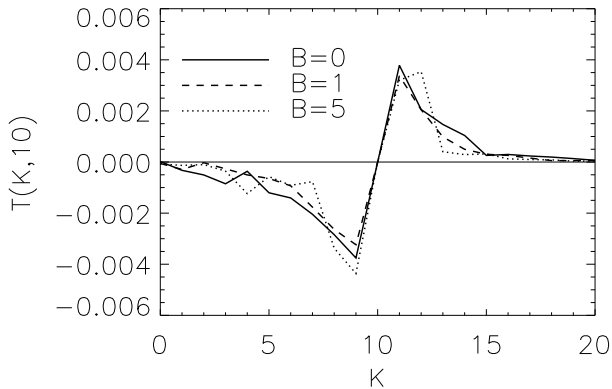}
\caption{The energy transfer function $\mathcal{T}^-(K,Q)$ for the perpendicular
cascade ({\it cylinders}) for $Q=10$ as a function of $K$, 
from the runs with 
$B=0$ (solid line), $B=1$ (dashed line) and $B=5$ (dotted line).
Amplitudes of $\mathcal{T}^-(K,Q)$ are normalized to have
the same order of magnitude than in the $B=0$ case.
\label{fig6}}
\end{figure}

\begin{figure}
\includegraphics[width=8cm]{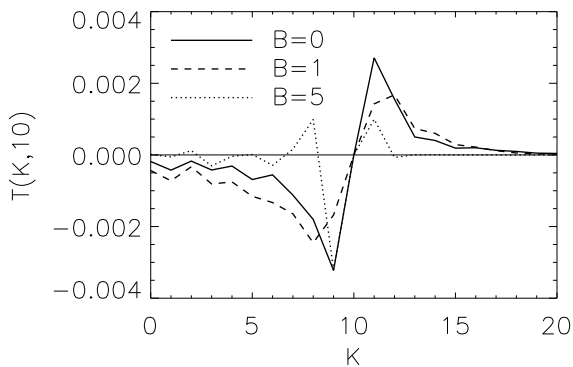}
\caption{The energy transfer function $\mathcal{T}^-(K,Q)$ for the parallel
cascade ({\it planes}) for  $Q=10$ as a function of $K$,
from the data at 
$B=0$ (solid line), $B=1$ (dashed line) and $B=5$ (dotted line).
Amplitudes of $\mathcal{T}^-(K,Q)$ are normalized to have
the same order of magnitude than in the $B=0$ case.
\label{fig7}}
\end{figure}

To get a better understanding of the $\mathcal{T}^\pm(K,Q)$ transfer functions, we look at 
a single wavenumber $Q$. Figure \ref{fig6} displays 
$\mathcal{T}^-(K,Q)$ for the perpendicular 
cascade (cylinders) at $Q=10$ as a function of $K$, whereas Figure \ref{fig7} shows it for
the parallel cascade (planes) at $Q=10$. 
To compare the results obtained from the different $B$ cases, 
the $\mathcal{T}^-(K,Q)$ amplitudes are normalized so that
all transfers are of the same order of magnitude.
Positive values of $\mathcal{T}^-(K,Q)$ imply that the shell $K$ receives
energy from the shell $Q=10$ (Fig. \ref{fig6}, perpendicular case) and 
 (Fig. \ref{fig7}, parallel case) while negative values of $\mathcal{T}^-(K,Q)$ 
mean that the shell $K$ gives energy to the shell $Q=10$.

For the perpendicular cascade, the shell $Q=10$ receives most energy from slightly smaller
wavenumbers than $K=10$ and it gives energy to slightly larger wavenumbers. 
This implies a locality in the energy transfer,
since it is mostly the nearby cylindrical shells that exchange energy. 
The parallel cascade presents a similar behavior; the shell $Q=10$ receives energy from slightly smaller
wavenumbers than $K=10$ and it gives energy to slightly larger wavenumbers. 
Note however that for the $B=5$ flow, there is also some trace of an inverse cascade 
(energy transfer from the wavenumber $Q=10$ to the wavenumber $K=8$).
This local behavior has also been found
in isotropic {($B=0$)} decaying MHD turbulence simulations \cite{Debliquy2005}.
Nevertheless, we need to note that in forced MHD turbulence
where the magnetic field is generated by dynamo action, strong nonlocal transfers also exist
\cite{Alexakis2005,Carati2006}. 
Whether these nonlocal transfers are present in the forced anisotropic
regime still needs further studies.

\subsection{Nonlinear interactions between ${\bf z}^+$ and ${\bf z}^-$}

\begin{figure}
\includegraphics[width=8cm]{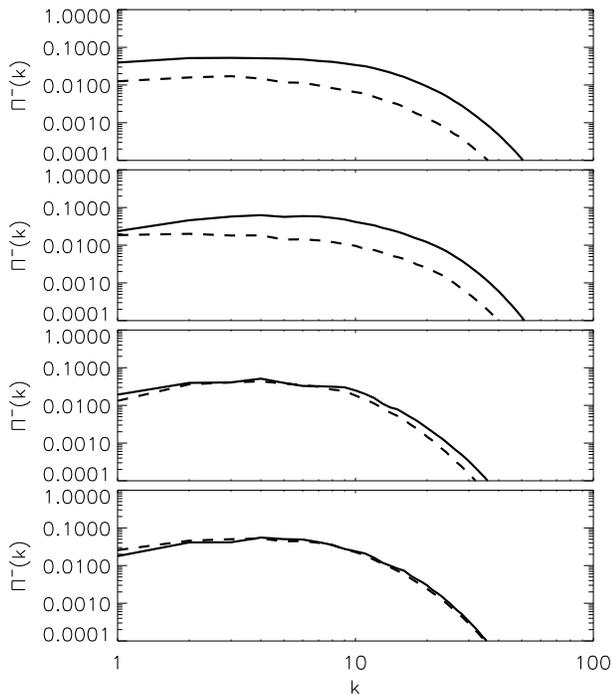}
\caption{The total energy flux $\Pi^{-}(K)$ (solid line)
across {\it cylinders} of radius $K$ together with the
partial flux $\Pi_{P=0}^{-}(K)$ (dashed line) 
for the four different values of $B$, from $B=0$ (top panel) up to
$B=15$ (bottom panel). 
\label{fig8}}
\end{figure}

The analysis of the energy transfer functions has thus shown that the energy cascades locally.
As a result, 
each Els\"asser field, ${\bf z}^+$ or ${\bf z}^-$, 
{ 
exchanges energy between waves traveling in the same direction of similar size}. 
Nonetheless, this does not mean that interactions 
among oppositely traveling waves are local. In the limit of very 
large intensities of the background magnetic field, where the weak turbulence theory is valid, 
the energy cascade is due to interactions with the modes in the plane at $k_\|=0$.
Therefore, modes with $k_\|\ge 1$ interact with modes $k_\|\ll1$ to cascade the energy.
To that respect, the interactions are nonlocal since short waves (large $k_\|$) interact
with long waves (small $k_\|$ ) to cascade the energy.
To investigate how close to the weak turbulence regime we are, 
we plot in Figure~\ref{fig8} the total energy flux $\Pi^{-}(K)$, defined in Eq.~(\ref{flux}), 
across cylinders together with the partial
flux $\Pi_{P=0}^{-}(K)$, defined in Eq.~(\ref{partialflux}) 
due to interactions with just $k_\|=0$ modes. 
As the strength of the uniform magnetic field is increased,
the flux due to the interactions with the $k_\|=0$ modes
become more and more dominant. 
In the $B=15$ flow, the global and partial fluxes  
{across cylinders}
become almost indistinguishable suggesting that 
interactions with the modes in the plane at $k_\|=0$  
are responsible for the energy cascade. {This means that the flow dynamics tends to be closer 
to a weak turbulence regime where the three-wave resonant interactions are dominating.}

\begin{figure}
\includegraphics[width=8cm]{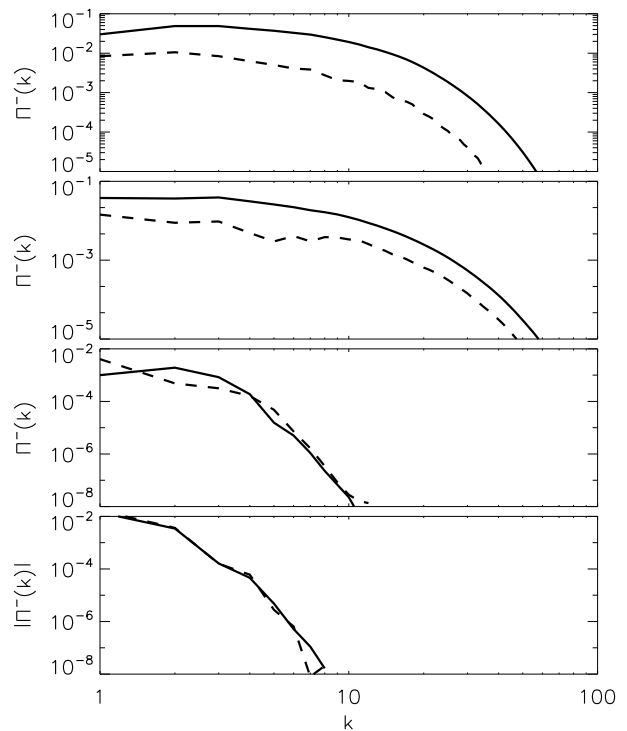}
\caption{The total energy flux $\Pi^{-}(K)$ (solid line)
across {\it planes} at $k_\|=K$ together with the
partial flux $\Pi_{P=1}^{-}(K)$ (dashed line)
for the four different values of $B$
from $B=0$ (top panel) up to $B=15$ (bottom panel).
\label{fig9}}
\end{figure}

{A different behavior is obtained for the parallel energy cascade.}
When a mode $\widehat{\bf z}^-({\bf k})$ interacts with a mode $\widehat{\bf z}^+({\bf p})$,
the $\widehat{\bf z}^-({\bf k})$ energy will move to the wavevector $\bf q$
so that the relation $\bf k+p+q=0$ holds. 
If however $\bf p$ belongs to the wavevector set
with $p_\|=0$, this relation then reads 
$k_\|+q_\|=0$ in the parallel direction, i.e. $|k_\||=|q_\||$. Therefore, the energy
remains in spectral planes located at the same distance from the origin. As a result, 
interactions with 
the $k_\|=0$ modes cannot contribute to the energy cascade in the parallel direction.
In this case, the closest modes to the $k_\|=0$ modes are the ones that gives most of the
energy flux.
Figure \ref{fig9} shows the total energy flux across planes and the partial
flux only due to interactions with the modes in the plane at $k_\|=1$ (the closest to
the $k_\|=0$ plane). 
As the amplitude of the ${\bf B}$-field is increased,
most of the parallel flux comes from modes with $k_\|=1$.
Here, we need to note that the flux in the parallel direction
is much noisier than the flux in the perpendicular direction
and that it often presents negative values 
(absolute values are plotted in the bottom panel of Fig. \ref{fig9}). 
A thorough analysis
of the parallel cascade would require to average many data outputs
which is not possible in the case of a freely decaying flows.
Such an analysis is left for future work.

\section{Conclusion and Discussion}

In this work we examine the energy cascade and
the interactions between different scales for freely decaying MHD flows
in the presence of a uniform magnetic field. 
Our analysis is based on data obtained from direct numerical 
simulations of the MHD equations with four different intensities of the applied magnetic field,
in an attempt to study the transition from strong to weak turbulence limit.
One clearly established result is that, 
as the strength of the uniform magnetic field is increased,
the energy spectrum becomes anisotropic with most of the energy concentrated in the 
small $k_\|$ wavenumbers, as already known \cite{Oughton94}.
It is further shown that the energy flux in the parallel direction (relatively to the uniform
magnetic field) is also strongly suppressed
when the guiding field in introduced.

To investigate the locality or nonlocality of the spectral interactions, 
we measure the transfer functions for the parallel and the perpendicular cascade.
The transfer functions in the parallel and perpendicular directions 
are found local whatever the strength of the  external magnetic field.
As a result, the coupling between modes that travel in the same direction
is local and the energy exchange occurs between similar size eddies. 
This behavior has been shown to hold in decaying isotropic MHD turbulence
simulations {(with $B=0$)} \cite{Debliquy2005}. 
However, in the presence of a mechanical forcing,
strong nonlocal interactions have been observed with a direct energy transfer 
from the forced scale to the inertial range scales \cite{Alexakis2005,Carati2006}. 
If this nonlocal behavior persists in the anisotropic case still needs 
further investigations.

The locality or nonlocality of the interactions between oppositely moving 
waves ($z^+$ and $z^-$), that do not exchange energy, is measured by means of  
partial fluxes in the parallel and the perpendicular directions due 
to the coupling in different spectral planes.
This coupling between oppositely propagating modes does not appear local.
As the amplitude of the applied magnetic field is increased,
most of the interactions occur with the $k_\|=0$ modes that are dominant in cascading
the energy.   
Most of the energy flux is thus in the perpendicular direction,
since the $k_\|=0$ modes do not contribute to the energy cascade in the parallel direction.
{Hence, the stronger magnetized flows tends to present a dynamics close to the weak turbulence
limit where the three-wave resonant interactions are responsible for the cascade process.}
This also partly explains the similar temporal evolution in the $B=5$ and $B=15$ regimes 
(see Figure \ref{fig1}) since, in both cases, most of the cascade 
is due to the $k_\|=0$ modes.

For the parallel cascade, the interactions are slightly different. As already said, this
is due to the inability of the $k_\|=0$ modes to cascade the energy in the parallel direction.
In that case, the modes with the smallest but not zero $k_\|$ ($k_\|\simeq1$) are the
ones responsible for the cascade.
This behavior is in qualitative agreement with the description of a recent phenomenological model
\cite{Alexakis2007}. However, the lack of resolution does not allow to pursue 
a quantitative comparison. 

Finally, we would like to emphasize that we analyze here numerical data 
of freely decaying MHD flows submitted to an external magnetic field
whose amplitude is varied, while all the other parameters are kept unchanged 
(periodic boundary conditions, unit magnetic Prandtl number, initial conditions and Reynolds number).
Thus, one should be cautious in any attempt to generalize the obtained results, e.g. 
forced turbulence could lead to different behaviors and should be studied separately.
The results could also be dependent on the kinetic Reynolds number as well on the magnetic Prandtl
number. 
{Furthermore the use of a refined spectral grid in the parallel direction, allowing 
the presence of more modes with $k_\|\simeq0$, could alter the energy cascade. 
}

\acknowledgments

This work was supported by INSU/PNST and /PCMI Programs and CNRS/GdR Dynamo. Computation 
time was provided by IDRIS (CNRS) Grand No. 070597, and SIGAMM mesocenter 
(OCA/University Nice-Sophia).


\begin{thebibliography}{}

\bibitem[Zeldovich, Ruzmaikin \& Sokoloff (1990)]{Zeldovich}
         Ya. B. Zeldovich, A.A. Ruzmaikin and D.D. Sokoloff
         ``Magnetic Fields in Astrophysics" 1990 Gordon \& Breach Science Pub.


\bibitem[Kolmogorov (1941)]{Kolmogorov1941}
         A.N. Kolmogorov,
         Dokl. Akad. Nauk SSSR {\bf 30},299 (1941)

\bibitem[Yousef, Rincon \& Schekochihin (2007)]{Yousef2007}
         T.A. Yousef, F. Rincon, and A.A. Schekochihin, J. Fluid Mech. {\bf 575}, 111 (2007)

\bibitem{Vermarev04}
         M.K. Verma, Phys. Reports, {\bf 401}, 229 (2004).

\bibitem[Verma, Ayyer, \& Chandra (2005)]{Verma2005}
         M.K. Verma, A. Ayyer, A.V. Chandra,
         Phys. Plasmas, {\bf 12}, 82307 (2005)

\bibitem[Alexakis, Mininni \& Pouquet (2005)]{Alexakis2005}
         A. Alexakis, P.D. Mininni and A. Pouquet,
         Phys. Rev. E {\bf 72}, 046301 (2005)

\bibitem[Mininni, Alexakis \& Pouquet (2005)]{Mininni2005}
         P.D. Mininni, A. Alexakis and A. Pouquet,
         2005, Phys. Rev. E 72, 046302

\bibitem[Carati, et al. (2006)]{Carati2006}
         D. Carati, O. Debliquy, B. Knaepen, B. Teaca and M.K. Verma
         J. Turb. {\bf 7}, 1, (2006)


\bibitem[Galtier et al.(2002)]{Galtier2002}
        S. Galtier, S.V. Nazarenko, A.C. Newell and A. Pouquet,
        \apj, {\bf 564}, L49 (2002)

\bibitem[Galtier et al.(2000)]{Galtier2000}
        S. Galtier, S.V. Nazarenko, A.C. Newell and A. Pouquet,
        J. Plasma Phys., {\bf 63}, 447 (2000)



\bibitem[Iroshnikov (1963)]{Iroshnikov1963}
         P. Iroshnikov,
         Soviet Astron., {\bf 7}, 566 (1963)


\bibitem[Kraichnan (1965)]{Kraichnan1965}
         R. Kraichnan, Phys. Fluids, {\bf 8}, 1385 (1965)

\bibitem[Goldreich \& Sridhar (1995)]{Goldreich1995}
         P. Goldreich and  S. Sridhar,
         \apj, {\bf 438}, 763 (1995)


\bibitem[Galtier et al.(2005)]{Galtier2005}
        S. Galtier, A. Pouquet, and A. Mangeney,
        Phys. Plasmas {\bf 12}, 092310 (2003)

\bibitem[Matthaeus \& Zhou (1989)]{Matthaeus1989}
         W.H. Matthaeus and Y. Zhou,
         Phys. Fluids B {\bf 1} 1929 (1989)


\bibitem[Zhou, Matthaeus \& Dmitruk (2004)]{Zhou2004}
         Y. Zhou, W.H. Matthaeus, P. Dmitruk
         Rev. Mod. Phys., {\bf 76}, 1015 (2004)

\bibitem[Bhattacharjee \& Ng (2001)]{Bhatta2001}
         A. Bhattacharjee and C.S. Ng,
         \apj, {\bf 548}, 318 (2001)

\bibitem[Alexakis (2007)]{Alexakis2007}
         A. Alexakis,
         { arXiv:0706.0816v1} (2007)


\bibitem[Biskamp \& M{\"u}ller (2000)]{Biskamp2000}
         D. Biskamp, and W.C. M{\"u}ller,
         Phys. Plasmas, {\bf 7}, 4889 (2000)

\bibitem[Cho \& Vishniac (2000)]{Cho2000}
         J. Cho  and E.T. Vishniac,
         \apj {\bf 539}, 273 (2000)

\bibitem[Dmitruk, Gomez, \& Matthaeus (2003)]{Dmitruk2003}
         P. Dmitruk, D.O. Gomez, and W.H. Matthaeus,
         Phys. Plasmas {\bf 10}, 3584 (2003)

\bibitem[Mason, Cattaneo \& Boldyrev]{Mason2007}
         J. Mason, F. Cattaneo and S. Boldyrev,
         astro-ph arXiv:0706.2003v1 (2007)

\bibitem[Maron \& Goldreich (2001)]{Maron2001}
         J. Maron and P. Goldreich,
         \apj {\bf 554}, 1175 (2001)

\bibitem[M{\"u}ller \& Grappin (2005)]{Muller2005}
         W.C. M{\"u}ller, and R. Grappin, Phys. Rev. Lett. {\bf 95}, 114502 (2005)

\bibitem[Ng \& Bhattacharjee (1996)]{Ng1996}
         C.S. Ng and A. Bhattacharjee,  \apj, {\bf 465}, 845 (1996)


\bibitem[Ng, Bhattacharjee, \& Germaschewski (2003)]{Ng2003}
         C.S. Ng,  A. Bhattacharjee, K. Germaschewski and S. Galtier,
         Phys. Plasmas, {\bf 10}, 1954 (2003)

\bibitem[Oughton, Priest, \&  Mattaheus (1994)]{Oughton94}
         S. Oughton, E.R. Priest and W.H. Mattaheus, J. Fluid Mech. {\bf 280}, 95 (1994).


\bibitem[Alexakis (2005)]{Alexakis2005b}
         A. Alexakis, P.D. Mininni and A. Pouquet,
         Phys. Rev. Lett. {\bf 95}, 264503 (2005)

\bibitem[Mininni, Alexakis \& Pouquet (2005)]{Mininni2006}
         P.D. Mininni, A. Alexakis and A. Pouquet,
         Phys. Rev E {\bf 74} 016303 (2006)

\bibitem[Milano et al.(2001)]{Milano}
         L.J. Milano, W.H. Matthaeus, and P. Dmitruk, D.C. Montgomery
         Phys. Plasmas {\bf 8}, 2673-2681 (2001).

\bibitem[Debliquy, Verma \& Carati (2005)]{Debliquy2005}
         O. Debliquy, M.K. Verma and D. Carati,
         Phys. Plasmas {\bf 12}, 042309 (2005)






\end{thebibliography}
\end{document}